\documentclass[12pt]{article}
\usepackage{amsmath,amsfonts,amssymb,a4,epsfig,color}

\newcommand{\R}{{\mathbb{R}}}
\newcommand{\C}{{\mathbb{C}}}

\newcommand{\N}{{\mathbb{N}}}

\def\ha{\frac{1}{2}}

\def\ra{\rightarrow}
\def\preuve{\begin{proof}} 
\def\ga{\alpha}

\def\gl{\lambda}
\def\go{\omega}

\newtheorem{defi}{Definition}

\newtheorem{lemm}{Lemma}
\newtheorem{prop}{Proposition}
\newtheorem{rem}{Remark}

\newtheorem{coro}{Corollary}
\newtheorem{theo}{Theorem}
\newtheorem{exem}{Example}[section]
\newenvironment{demo}{\noindent {\it Proof.--}
      \begin{quotation}\noindent}{\end{quotation}\hfill$\square $}

\DeclareMathOperator{\ind}{ind}

\begin{document}

\title{Magnetic interpretation of the nodal defect on graphs}
\author{Yves Colin de Verdi\`ere \footnote{Institut Fourier,
 Unit{\'e} mixte
 de recherche CNRS-UJF 5582,
 BP 74, 38402-Saint Martin d'H\`eres Cedex (France);
yves.colin-de-verdiere@ujf-grenoble.fr}
\footnote{Thanks to Roland Bacher, Gregory Berkolaiko 
  and Fran\c{c}oise Truc for carefully
  checking the manuscript.}}


\maketitle
\begin{abstract}
\begin{color}{blue}In this note, we  present a natural proof 
 of a recent and surprising  result of Gregory Berkolaiko interpreting the 
Courant nodal defect as a Morse index. This proof is inspired
by a nice paper of Miroslav Fiedler published in 1975.
\end{color}
\end{abstract}

\section{Introduction}
The ``nodal defect'' of an eigenfunction of a Schr\"odinger  operator 
is closely related to the difference between the upper bound
on the number of nodal domains given by Courant's Theorem
and the number of nodal domains.
In the recent paper \cite{Berk1}, Gregory Berkolaiko proves
 a nice formula for
the  nodal defect  of an eigenfunction of a Schr\"odinger  operator 
on a finite graph in terms of the Morse index of the corresponding
eigenvalue as a function of  a magnetic deformation of the operator.
His proof remains mysterious and rather indirect. In order  to get
a better understanding in view of possible generalizations, it 
is desirable to have a more direct approach. This is what we do here. 

After reviewing our notation, we summarize the main result and give an
informal description of the proof in Section~\ref{sec:statement}.  The
proof itself is implemented in Sections~\ref{sec:qf} and
\ref{sec:hess} with an alternative view provided in
Appendix~\ref{sec:pedestrian}.  The continuous Schr\"odinger operator
on a circle is considered in Appendix~\ref{sec:Hills} and various
special cases and further ideas are explored in other Appendices.

\section{Notation}
\label{sec:notation}

Let $G=(X,E)$ be a finite connected graph where $X$ is the set of
vertices and $E$ the set of  unoriented  edges. We denote by
$\{ x,y \}$ the edge linking the vertices $x$ and $y$. 
 We  denote by $\vec{E}$ the set of oriented edges and 
 by  $[x,y]$ the edge from
$x$ to $y$;
  the set  $\vec{E}$  is  a 2-fold cover of $E$.
A 1-form $\ga $ on $G$ is a map  $\vec{E}\ra \R $ such that
$\ga([y,x])=-\ga([x,y])$
for all  $\{x,y \}\in E$. 
 We denote by $\Omega ^1(G)$ the vector space
of dimension $\# E$ of 1-forms on $G$.
The operator $d:\R^X  \ra \Omega ^1(G) $ is defined by
$df([ x,y ])=f(y)-f(x)$.
 If $Q$ is a non-degenerate, not necessarily positive,
quadratic form 
on $\Omega ^1(G)$, we denote by $d^\star $ the adjoint of $d$ where
$\R^X$ carries  the canonical Euclidean structure
and $\Omega ^1(G)$ is equipped with the symmetric  inner  product $\hat{Q}$ 
associated to  $Q$.
 We have $\dim {\ker d^\star}=\beta$ where
 $\beta =1+\# E -\# X$ is the dimension
of the space of cycles of $G$.
We will show later that, in our context, we have the Hodge decomposition 
$\Omega^1(G)=d\R^X \oplus {\ker d^\star} $ where both spaces are
 $\hat{Q}$-orthogonal.

Following \cite{YCdV1}, we denote by ${\cal O}_G $ the set
of   $X \times X$ real symmetric matrices $H$ which satisfy
$h_{x,y}<0 $ if $\{ x, y\} \in E$
and $h_{x,y}=0 $ if $\{ x, y\} \notin E$ and $x\ne y$. Note that the
diagonal
entries of $H$ are arbitrary.
An element $H$ of  ${\cal O}_G $ is called a {\it Schr\"odinger operator}
on the graph $G$.
It will be useful to write
the quadratic form associated to $H$ as
\[ q_1(f)=-\sum_{\{ x,y\} \in E}
 h_{x,y}(f(x)-f(y))^2 + \sum_{x\in X} V_x f(x)^2 ~,\]
with $V_x=h_{x,x}+\sum_{y\sim x}h_{x,y}$. 
A {\it magnetic field} on $G$  is a map $B :\vec{E} \ra U(1)$
defined  by $B([x,y ])= e^{i \ga _{x,y}}$
where  $[x,y] \mapsto  \ga _{x,y}$ is a 1-form on $G$.
We denote by ${\cal B}_G=e^{i\Omega ^1(G)}$ 
the manifold of magnetic fields on $G$.
The magnetic Schr\"odinger operator $H_B$ associated to $H\in {\cal O_G}$
and $B=e^{i\ga}$
is defined by the quadratic  form
\[ q_B(f)=-\ha \sum_{[x,y]\in \vec{E}} h_{x,y}| f(x)-e^{i\ga_{x,y}}f(y)|^2+
\sum_{x\in X} V_x |f(x)|^2  \]
 associated to a Hermitian form on $\C^X$.
More explicitly, if $f\in \C^X$, 
\begin{equation}\label{equ:magn}
 Hf(x)=h_{x,x}f(x) + \sum_{y\sim x} h_{x,y}e^{i\ga _{x,y}}f(y)~.
\end{equation}

We  fix $H$ and we denote by 
\[ \gl_1 (B) \leq \gl_2 (B) \leq \cdots \leq \gl_n(B) \leq \cdots
\leq \gl_{\# X}(B) \]
the eigenvalues of $H_B$. 
It will be important to notice that $\gl_n (\bar{B})=\gl_n(B)$.
Moreover, we have a gauge invariance:
the operators 
$H_B$ and $H_{B'}$  with 
$\ga ' =\ga + df $ for some $f\in \R^X$  are unitarily equivalent. Hence 
they have the same eigenvalues.
This  implies that, if $\Omega ^1(G)=d\R^X \oplus {\ker d^\star} $
(this is not always the case because $Q$ is not positive),
it is enough to consider
$1-$forms in the subspace $\ker d^\star $ of $\Omega ^1(G)$ when
studying
the map  ${\Lambda _n}:B \ra \gl_n(B)$. This holds in particular
 for investigations
 concerning the Hessian and the Morse index.

\section{Statement of Berkolaiko's magnetic Theorem}
\label{sec:statement}

Before stating the main result, we recall the
\begin{defi}
\label{def:index}
The {\rm Morse index} $j(q)\in \N \cup \{ +\infty \} $
  of a quadratic form $q$ on a real  vector space $E$
is  defined by 
$j(q)=\sup_{F  } \dim F$ where $F$ is a subspace of $E$ so that
$q_{|F\setminus 0}$ is $<0$. The  {\rm nullity} of $q$ is the dimension 
of the kernel of $q$.

The {\rm Morse index} of a smooth real-valued  function $f$ defined
on a smooth manifold $M$  at a  {\rm critical point} $x_0 \in M$
 (i.e. a point satisfying
$df(x_0)=0$) is the Morse index of the Hessian of $f$, which is a
canonically  defined quadratic form on the tangent space $T_{x_0}M$.
The critical point $x_0$ is called {\rm non-degenerate} if the previous
Hessian is  non-degenerate.
The {\rm nullity} of the critical point $x_0$ of $f$  is the nullity of the
Hessian
of $f$ at the point $x_0$. 
\end{defi}

The aim of this note is to prove the following nice results due to 
Berkolaiko \cite{Berk2,Berk1}:
\begin{theo} \label{theo:main}
 Let $G=(X,E)$ be a finite connected graph  
 and $\beta $ 
the dimension of the space of cycles of $G$.
We suppose that the $n$-th eigenvalue $\lambda _n$ of  $H\in
{\cal O}_G$ 
is simple. We assume moreover 
 that     an associated non-zero eigenfunction $\phi _n$
satisfies  $\phi_n (x)\ne 0$ for all $x\in X$.
Then, the number $\nu $ of edges along which $\phi_n$ changes sign
satisfies
$n-1 \leq \nu \leq n-1 +\beta  $.

Moreover ${\Lambda _n}:B \ra \gl_n (B)$ is smooth at $B\equiv  1$ which 
is a critical point of ${\Lambda _n}$ and  
 the {\rm nodal defect}, $\delta _n =\nu -(n-1)$ is
the Morse index  of ${\Lambda _n} $ at
that point.
If $M$ is the manifold of dimension $\beta$ 
 of magnetic fields on $G$ modulo
the gauge transforms, the function $[B] \ra \Lambda _n (B)$ has
$[B=1]$ as a non-degenerate critical point. 
\end{theo}
\begin {rem} The previous results can be extended by replacing
the critical point $B\equiv 1$ by $B_{x,y}=\pm 1$ for all edges
$\{x,y\} \in E$.
The number  $\nu $ is then 
 the number of edges $\{x,y\}\in E$  satisfying
$B_{x,y}\phi_n(x)\phi_n(y)<0$ where $\phi_n$ is the corresponding eigenfunction.
\end{rem}
\begin {rem}The assumptions on $H$ are satisfied for $H$
in an open dense subset of
${\cal O}_G $. \end{rem}
The upper bound of $\nu$ in the first part
of Theorem \ref{theo:main}  is related to Courant nodal Theorem (see
\cite{Courant-Hilbert}
Section  VI.6)
  as follows: a nodal domain on a graph for the eigenfunction
$\phi_n$  is a connected
 component of the sub-graph $G'$
of $G$ obtained by removing the edges along which $\phi_n$
changes sign.
Denoting  by $\mu $ the number of nodal domains of $\phi_n$, the Courant
Theorem for graphs
(see \cite{YCdV1}, Theorem 2.4)  asserts that $\mu \leq n $;
 using Euler formula 
for the graph $G'$   and because $\mu= b_0 (G') $, the number of
connected
components of the graph $G'$, we get also a lower bound (see \cite{Berk2}):
\begin{coro}
Under the assumptions of  Theorem \ref{theo:main}, we have
$ n-\beta \leq \mu \leq n~.$
\end{coro}

{\bf Important warning:}
{\it Without loss of generality, we can and WILL assume
in the rest of this note  that $\gl_n=\Lambda_n(1)=0$. This implies that
the Morse index of $q_1$ is $n-1$.  }

In the course of the proof we will use a special choice of gauge in
which we can compute the Hessian explicitly.  More precisely,
according to the classical perturbation formulae,
\begin{equation*}
  \ddot\lambda = (\phi, \ddot{H}\phi) 
  + 2(\dot{H} \phi, \dot\phi),
\end{equation*}
where we assumed that $\lambda$ is at a critical point: $\dot\lambda =
0$.  The first term is easy to calculate explicitly; for perturbation
in the direction of the 1-form $\go$ it is
\begin{equation} \label{equ:axy}
  Q(\go)=\ha \sum _{\vec{E}}  a_{x,y} \go ([x,y])^2 ~{\rm with }~ 
  a_{x,y}=-h_{x,y}\phi_n(x)\phi_n(y)=a_{y,x}~.
\end{equation}
Considered as a quadratic form in $\go$, $Q$ is already in the diagonal
form.  Its index is clearly the number of negative values among
$\left\{-h_{x,y}\phi_n(x)\phi_n(y)\right\}$, or, in other words, the
number $\nu$ of edges where $\phi_n$ changes sign! 

We will present an explicit choice of gauge in which the second term
vanishes.  The condition for this is $\dot{H} \phi=0$ which, after
explicit calculation, can be interpreted as $\go \in \ker d^\star$,
where $d^\star$ is the conjugate of $d$ with respect to the inner
product induced by \eqref{equ:axy}.

Finally, we observe that the index of $Q(\go)$ has been computed to be
$\nu$ in the whole of $\Omega^1(G)$, whereas we should be restricting
ourselves to our chosen gauge, $\go \in \ker d^\star$.  We will show that this
restriction reduces the index precisely by $n-1$.  Indeed, the
splitting $\Omega^1(G) = d\R^X \oplus {\ker d^\star}$ is orthogonal
with respect to the form $Q$, therefore
\begin{equation*}
  \ind(Q) = \ind\left(Q|_{d\R^X}\right) + \ind\left(Q|_{\ker
      d^\star}\right).
\end{equation*}
We establish that ${\rm ind }(Q|_{d\R^X})=n-1$ by relating the form
$Q$ on $d\R^X$ to the quadratic form $q_1$ around the point $\phi_n$.

\section{The quadratic  form $Q$}
\label{sec:qf}

\begin{lemm}
The set of forms $f\ra (f(x)-f(y))^2$ where $\{x,y\}\in {\cal
  P}_2(X)$,
the set of    subsets with
two  elements of $X$,  and $f
\ra f(x)^2$ with $ x\in X$
is a basis of the set of quadratic forms on $\R^X$.
\end{lemm}
\begin{defi}
 A quadratic form $q$ on $\R^X $  is said of Laplace type if
  $~\forall f \in \R^X,~\hat{q} (1,f)\equiv 0$
where $\hat{q}$ is the symmetric bi-linear form associated to $q$.
\end{defi}
\begin{lemm} \label{lemm:lapl}
 The set of forms  $f\ra (f(x)-f(y))^2, ~\{x,y\} \in {\cal P}_2(X)$
is  a basis of the space of  quadratic forms  of Laplace type.
\end{lemm}

The form $\tilde{q}_1:f\ra q_1(\phi_n f)$, where $\phi_n f$ is the
point-wise
product of $\phi_n$ and $f$,  
is of Laplace type because
\[ \widehat{\tilde{q}_1}(1,g)=\langle H\phi_n |\phi_ng \rangle =
\langle 0 |\phi_ng \rangle~.\]
 Hence $\widehat{\tilde{q}_1}(1,g)=0$.

Moreover, $\tilde{q}_1(f) = Q(df)$.  Indeed, because of Lemma
\ref{lemm:lapl},
 it is
enough to compare the coefficients of the basis forms $f \ra (f(x) -
f(y))^2$.  The form $f\ra Q(df)$ is already expanded in this basis.  To
find the coefficient for the form $f\ra \tilde{q}_1(f)$, we observe that
(because we know it is of Laplace type) the coefficient in question is
minus 
the coefficient in front of the term $f(x)f(y)$, divided by two.  This
evaluates to $a_{x,y}$ (see equation (\ref{equ:axy})).

In fact, we will need to use
$ \hat{Q}(df,dg)=\langle H(\phi_n f) |\phi_n g \rangle $.
\begin{lemm} The Morse index of $Q_{|d\R^X}$ is equal
to $n-1$.
\end{lemm}
It is a general  fact that the Morse index of the quadratic form
$f\ra Q(Af)$ is the same as the Morse index of the restriction of
$Q$ to the image of $A$.
Hence, the Morse index of  $Q_{|d\R^X}$ is the Morse index 
of $\tilde{q}_1$ on $\R^X$. 
Because $f\ra \phi_n f $ is a linear isomorphism, this index  is equal
 to  the index of
$q_1$ by Sylvester Theorem. 
 Since $\gl_n=0$,
 the index of $q_1$ is  $n-1$ by elementary spectral theory.

\begin{lemm}\label{lemm:Hodge}
Let us denote by $d^\star $ the adjoint of $d$ where
$\R^X$ is equipped with the canonical Euclidean structure and
$\Omega ^1 (G)$ with the inner product associated to $Q$.
The space $\Omega ^1 (G)$ splits as
\[ \Omega ^1 (G)=d\R^X \oplus {\ker d^\star} \]
 (Hodge type splitting), 
and this decomposition is $Q$-orthogonal.
\end{lemm}
More explicitly $d^\star $ is given 
by 
\[ ~d^\star \go (x)=\sum _{y\sim x} a_{x,y}\go([y,x])~.\]

If $\go =df $ satisfies $d^\star \go=0$, we have 
$d^\star d f=0 $. Hence
$\hat{Q} (df, dg)=0$ for all $g$  and
$\langle H(\phi_n f )|\phi_n g)\rangle =0$.
Because $\gl_n$ is of multiplicity $1$, this implies that
$f$ is constant and hence $df=0$.
So $d\R^X \cap {\ker d^\star}=\{ 0 \}$ and the conclusions follow.

At this point, we know that the nodal defect is the Morse index
of the restriction of $Q$ to the space ${\ker d^\star}$ of dimension $\beta $.
The first part of the Theorem follows.

\section{The magnetic Hessian}
\label{sec:hess}

We need one more fact to complete the proof: to 
identify the Hessian of ${\Lambda _n}$ on $e^{i{\ker d^\star}}$
at $B\equiv 1$ with   
the restriction of $Q$ to ${\ker d^\star}$.

Let us denote by $S\subset  \C^X$ the set of unit vectors $f$
 normalized so that $f(x_0)$  is real and $f(x_0)>0$
 where $x_0$ is chosen in $ X$.
\begin{lemm}
The point $B\equiv 1$ is a critical point of ${\Lambda _n}$.
If $\phi_n(B)\in S $ is the eigenfunction of $H_B$ corresponding to
the eigenvalue $\gl_n (B)$, the differential of $B \ra \phi_n (B)$
vanishes
at $B\equiv 1$ on ${\ker d^\star}$. 
\end{lemm}
The first property comes from
 the fact that ${\Lambda _n}(\bar{B})={\Lambda _n}(B)$.
We can compute, for any variation $e^{it\ga }$, $t$ close to $0$,  
of $B\equiv 1$, 
$\dot{H}_B\phi_n + H\dot{\phi}_n =0$.
The condition $d^\star \ga =0$  can be written as 
 $\sum _{y\sim x}h_{x,y}\phi_n(y) \ga _{x,y}=0$ for all  $ x\in X$.
From Equation (\ref{equ:magn}), this is equivalent to  
$\dot{H}_B\phi_n =0$.
Hence $H(\dot{\phi_n})=0$
and $\dot{\phi_n}=c\phi_n$ since  $\gl_n $
is simple.  From the normalization  $\| \phi_n(B)\| =1$, we get
$c\in i\R $ and, since $\dot{\phi_n } (x_0)\in \R $, the number  $c$ 
is real.
We deduce that $\dot{\phi_n}=0$.

\begin{lemm}\label{lemm:hess} The function
 $F:S \times e^{i{\ker d^\star}} \ra \R $
defined by $F(f,e^{i\ga} )= \langle H_{e^{i\ga}} f |f \rangle $
admits $(\phi_n ,0)$ as a critical point 
and the Hessian of $({\Lambda _n})_{|e^{i{\ker d^\star}}} $
 at the  point $B\equiv 1$  is
the form $Q$.
\end{lemm}
The differential of $F$ with respect to $f$ vanishes because
$f$ is an eigenfunction of $H$.
The differential with respect to ${\ker d^\star}$ vanishes, because
$F(f,e^{i\ga} )=F(f,e^{-i\ga} )$. The Hessian
of $F$ at $(\phi_n,0)$  is well defined.
Because the differential at $B=1$ of 
$B\ra \phi_n (B)$ vanishes on $e^{i{\ker d^\star}}$,
 the Hessians
of ${\Lambda _n}:B\ra F(\phi_n (B),B)$ and $M_n:B\ra F(\phi_n(1) ,B)$ agree.
A simple calculation of the Hessian of $M_n$ gives the result:
\[ M_n (e^{i\ga})=-\ha \sum _{[x,y]\in\vec{E}}
h_{x,y}|  \phi_n (x)-e^{i\ga _{x,y}}\phi_n (y)|^2 +\sum
_{x\in X}V_x |\phi_n(x)|^2=\]
\[ ~ -\sum _{[x,y]\in {E}}h_{x,y}
\left( \phi_n(x)^2 + \phi_n(y)^2 -2\cos \ga _{x,y}\phi_n(x)\phi_n(y)\right)+\sum
_{x\in X}V_x |\phi_n(x)|^2 .\]
Computing the second derivative with respect to $\alpha $ at
$\ga =0$ gives
${\rm Hessian}(M_n)=Q(\alpha )$.

\appendix
\section{A pedestrian approach 
to the calculus of the Hessian of ${\Lambda _n}$
in Section \ref{sec:hess}}
\label{sec:pedestrian}

We will derive a direct approach to the calculus of the second 
derivative of an eigenvalue which could be used directly 
in the proof of Lemma \ref{lemm:hess}.
Let $t \ra A(t)$ be a $C^2$ curve defined near  $t=0$ in the space 
of Hermitian matrices
on a finite dimensional Hilbert space
$({\cal H},\langle .|. \rangle )$.
Let us assume that $\gl(0)$ is an eigenvalue of $A(0)$ of multiplicity 
one with a normalized eigenvector $\phi (0)$.
Then, for $t$ close to $0$, $A(t)$ has a simple eigenvalue
$\gl(t)$ of multiplicity one which is a $C^2$ function of $t$.
We can choose an associated eigenfunction $\phi(t)$ which is $C^2$
with respect to $t$. 
The following assertions give the values of the first and second
derivatives
of $\gl(t)$ at $t=0$:
\begin{prop}\label{prop:deriv}
Under the previous assumptions, we have
\[ \gl'(0)= \langle A'(0) \phi(0) | \phi (0) \rangle ~,\]
If  $\gl'(0)=0$, we have
\[ \gl'' (0)=\langle A''(0) \phi(0) | \phi(0) \rangle +
2 \langle \phi'(0)|
A'(0 )\phi (0)\rangle ~,\]
where $\phi'(0)$ is any solution of  $(A(0)-\gl(0))\phi'(0)=-A'(0)\phi(0)$. 

In particular, if $A'(0 )\phi(0)=0$, 
\[ \gl'' (0)=\langle A''(0) \phi(0) | \phi(0) \rangle~.\] 
\end{prop}
\begin{demo}
We start with $(A(t)-\gl(t))\phi(t)=0 $ where $\phi(t)$
is  an eigenfunction of $A(t)$
which depends in a $C^2$ way of $t$. 
Taking the first derivative, we get
\begin{equation}\label{equ:first}
(A'(t)-\gl'(t))\phi(t)+(A(t)-\gl(t))\phi'(t)=0~.\end{equation}
Putting $t=0$ and taking  the scalar product with $\phi(0)$,
we get the formula for $\gl'(0)$.
Similarly, the $t$-derivative of Equation (\ref{equ:first})
is 
 \begin{equation}\label{equ:second}
(A''(t)-\gl''(t))\phi(t)+2(A'(t)-\gl'(t))\phi'(t)+
(A(t)-\gl(t))\phi''(t)=0~.\end{equation}
Pouting $t=0$, taking  the scalar product with $\phi(0)$
and using $\gl'(0)=0$, we get
the result.
\end{demo}

We can apply this to $A(t):=H_{e^{it\ga}}$ with $\ga \in {\ker d^\star} $
in order to get the Hessian of ${\Lambda _n}$ in Section \ref{sec:hess}.
The condition $A'(0)\phi(0)=0$ is exactly $d^\star \ga =0$!

\section{Hill's operators}
\label{sec:Hills}

In this Appendix, we will describe the case of a Schr\"odinger 
operator on the circle, also called the Hill's operator.
This is the simplest continuous case, but it may be useful to do
it with some details in order to try to extend the method to
higher dimensional manifolds.

\subsection*{Eigenvalues and discriminant}
The Hill's operator is 
\[ H=-\frac{d^2}{dx^2}+q(x)\]
where $q:\R \ra \R  $ is a smooth, $1$-periodic,  function.
The spectral theory of Hill's operators has been well studied; 
in particular, 
 the inverse spectral theory 
for this operator allows to solve
non-linear evolution equations, like the Korteweg-de Vries one.
A presentation of the properties of Hill's operators
is given  in \cite{M-W}. 

The following facts are known:
\begin{theo}
If we denote by $\gl^\pm _j,~j=1,\cdots $
the spectra of $H$ acting on periodic (resp anti-periodic) functions
of period $1$, we have the inequalities
\[ \gl_1^+ <\gl_1^-\leq \gl_2^-< \gl_2^+\leq \gl_3^+< \cdots \]
and the spectrum of $H$ on $L^2(\R)$ is then union of intervals,
called the {\rm bands}, 
\[ [\gl_1^+ ,\gl_1^-]\cup  [\gl_2^-, \gl_2^+]\cup [
\gl_3^+,\gl_3^-]\cup
\cdots ~.\]
\end{theo}
These statements are linked to the properties of the {\it discriminant}
$\Delta (\lambda )$: if 
$y_1(x,\lambda )$ and $y_2(x,\lambda )$
are the {\it normalized solutions} of
$(H-\gl)y=0$ whose Cauchy data are
$y_1(0,\lambda )=1,~y_1'(0,\lambda )=0,~y_2(0,\lambda )=0,~
y_2'(0,\lambda )=1$, the discriminant $\Delta $ is the   entire function
  given by $\Delta( \gl):=y_1(1,\lambda)+y_2'(1,\lambda)$.
The spectrum of $H$ on $L^2(\R)$ is the set of real
$\lambda $'s so that $|\Delta (\lambda )|\leq 2$.
The periodic (resp. anti-periodic) spectra are given
by $\Delta (\lambda )=2$ (resp.  $\Delta (\lambda )=-2$).
The function $\Delta (\lambda )-2 $ is a regularization
of $\prod_{n=1}^\infty  (\lambda -\gl_n^+)$ in the spirit
of \cite{YCdV2}.
It is proved in \cite{M-W}, Section II, that, 
if $\gl_n^+$ is simple, $\Delta '(\gl_n^+ )\ne 0$
and the sign of this derivative is that of $(-1)^n$.

\subsection*{Magnetic fields}

We will assume that $\gl_n^+ $ is equal to $0$ and  is a simple eigenvalue of
$H$ acting on $1$-periodic functions. 
Up to gauge transform, every magnetic potential on the circle
is a constant $\ga$.
The bands are linked to the addition of a magnetic field as follows:
the $n$-th band is the image of the circle
$U=\{ e^{i\ga }|\ga \in \R \}$ by the map
$\Lambda _n $ where $\Lambda _n(e^{i\ga})$
is the $n$-th eigenvalue of $H_\ga $ which is $H$ acting on functions
$f$ so that $f(x+1)=e^{i\ga}f(x)$.
In particular,  if $n$ is even, $\gl_n^+$ is a maximum
of $\Lambda _n $ while  if $n$ is odd, $\gl_n^+$ is a minimum
of $\Lambda _n $.
This fits with Berkolaiko's formula because the (even!) number of zeros
of the corresponding periodic eigenfunction $\phi_n$
is $n=(n-1)+1$ if $n$ is even and $n-1=(n-1)+0$ if $n$ is odd (see
\cite{M-W} Theorem 2.14).
In this appendix, we will use the general formula
for the second derivative in order to reprove
this result and to show that the critical points are non-degenerate. 

A  direct computation of $d^2\Lambda _n/d\ga^2 (0)$  using the
discriminant
works  as follows:
the spectrum of $H_\ga $ is given by
$\Delta ^{-1}(2 \cos \ga)$. 
Near $\gl =\gl_n^+$, we have
$2+ \Delta '(\gl_n^+)(\gl_n (\ga) -\gl_n^+) \sim 2\cos \ga$.
This gives $\gl_n  (\ga) \sim \gl_n^+- \ga ^2/ \Delta '(\gl_n^+)$,
hence the Morse index of $\Lambda _n $
at $\ga =0$ is $0$ if $n$ is odd and $1$ is $n$ is even. 

\subsection*{A direct calculation of the Hessian}
We will denote with a ``dot'' the derivatives w.r. to $\ga$
and by a ``prime'' the derivatives w.r. to $x$.
The operator $H_\ga $ is unitarily equivalent to
$K_\ga =e^{-i\ga x}He^{i\ga x} $
acting on $1$-periodic functions.
We have
\[ K_\ga =H-2i\ga \frac{d}{dx}+\ga ^2 ~.\]
The derivatives of $K_\ga $ w.r. to $\ga $
at $\ga =0$ are 
$\dot{K}=-2i \frac{d}{dx}$ and $\ddot{K}=2$.
Applying Proposition  \ref{prop:deriv}
and denoting by $\phi_n $ a corresponding normalized 
eigenfunction, we get
\[ \ddot{\Lambda} _n(0)=2 +4i\int_0^1 \dot{\phi}_n (x)\phi_n'(x)dx ~.\]
 Moreover $H\dot{\phi}_n (x)=-\dot{K} \phi_n= 2i \phi_n'(x)$.

Let us denote by $\psi $ the function $y_1(.,0)$.
Then, using the method of ``variation of parameters'' (i.e.
making the Ansatz 
$ \dot{\phi}_n (x)=C_1(x)\psi(x)+ C_2(x) \phi_n (x)$
with $C_1'(x)\psi(x)+ C_2'(x) \phi_n (x)=0$), we get
\begin{equation}\label{equ:phin}
 \dot{\phi}_n (x)=-ix\phi_n(x)+k \psi(x)+ C\phi_n (x)~,
\end{equation}  
where the constant $k$ is chosen so that  $\dot{\phi}_n (x)$
is periodic and $C$ is an arbitrary constant which can be fixed by a
normalization of $\phi_n$.
 We can always assume that $\phi_n(0)=\phi_n(1)=0$
by shifting the origin of $\R$ to some zero of $\phi_n$.
Using the wronskian, we see that  $\dot{\phi}_n (1)=\dot{\phi}_n (0)$.
We have to check the derivatives:
$k\psi'(1)-i(\phi_n(1)+ \phi'_n(1))=k\psi'(0)-i\phi_n(0)$
or
$k\psi'(1)=i\phi'_n(0)$. This gives, using Equation (\ref{equ:phin}), 
\[ \dot{\phi}_n (x)=-ix\phi_n(x)+i\frac{\phi'_n(0)}{\psi'(1)} \psi(x)+
C\phi_n(x)~.\] 

We get 
\[ \ddot{\Lambda }_n (0)=2+4i\int_0^1[-ix \phi_n(x)+k\psi(x)+
C\phi_n(x)]\phi_n'(x)dx
~. \]
By integration by parts, 
we have $\int_0^1 2 x \phi_n(x)\phi_n'(x)dx=-\int_0^1 \phi_n(x)^2 dx
=-1$.
Moreover, again by integration by parts, 
$\int_0^1 \psi(x)\phi_n'(x)dx=-\int_0^1 \psi'(x)\phi_n(x)dx $
and, 
since the Wronskian $\psi  \phi_n' - \psi' \phi_n$ is
constant and  $\equiv \phi_n'(0)$,
 $\int_0^1 \psi(x)\phi_n'(x)dx=\ha \phi_n'(0)$.
We get 
\[\ddot{ \Lambda }_n (0)=-2\phi_n'(0)^2/\psi'(1)~.\]
Moreover, it follows from Equation (2.13), page 16 in \cite{M-W}
and the fact that $\phi_n=\phi'_n(0 )y_2$, 
that this is exactly $-2/\Delta '(\gl_n ^+)$.

\section{The case where the
eigenfunction vanishes at some vertex}
\label{sec:vanishing}

In this Appendix,  we take $H\in  {\cal O}_G$
and assume that $\gl_n=0$ is  non-degenerate eigenvalue of $H$
with a normalized eigenfunction $\phi$.
We have the 
\begin{prop}\label{prop:var}
Let us assume  that, for all vertices $x$ satisfying 
$\phi(x)=0$,
 there
exists a vertex 
$y\sim x $ so that $\phi(y)\ne 0$.
Then, for any $\psi \in \R ^X $ orthogonal to $\phi$, there exists
a smooth deformation $H_t \in {\cal O}_G $ of $H$ so that $\dot{\phi}=\psi $.
\end{prop}
It  is enough to check that the space of $\dot{H}\phi $
is $\R^X$ and to use the first variation formulae given in Appendix A. 
\begin{theo}
Let us assume that the function $\phi $
vanishes at the unique vertex  $x_0$.
Then, the nullity of the Hessian of the ''magnetic variation''
of $H$ is at least $|n_+ -n_-|$ where
$n_\pm $ is the number of vertices $x\sim x_0$ so that
$\pm \phi (x)>0$.
\end{theo}
\begin{demo}
Choose a smooth variation $H_t$ of $H$ so that
$\dot{\phi}(x_0)=1$. Let  $\nu $ be the number of
sign changes of $\phi $ away of $x_0$.
Then, for $t>0$ small enough, the number of sign change
of $\phi_t$ is  $\nu +n_-$ while, for  $t<0$ small enough,
it is  $\nu + n_+ $.
We see from Theorem \ref{theo:main} that the magnetic Morse
index is $\nu +n_- -(n-1)$ for $t>0$ and
 $\nu +n_+ -(n-1)$.
The discontinuity of the Morse index at $t=0$ is $|n_+ -n_-|$. 
This gives the lower bond on the nullity.
 \end{demo}

\begin{coro} If $|n_+ -n_-|  > \beta $, the eigenvalue
$0$ is degenerate.
\end{coro}

Let us remark that this lower bound is not always sharp.
In the following example, we have $n_+=n_-$, $\beta =2$
and the nullity of the Hessian is $2$.
\begin{exem}
The graph $G$ is made of 2 cycles of length 3 with a common vertex.
The matrix of $H$ is chosen as follows:
\[ [H]=-\left(\begin{array}{ccccc}1&1&1&0&0 \\
                                  1&1&2&0&0 \\
                                  1&2&1&1&2 \\
                                  0&0&1&1&1 \\
                                  0&0&2&1&1 
\end{array}\right)\]
Using the fact that the graph has a symmetry of order 2 exchanging
the 2 cycles, one can split $\R^X $ and the matrix $H$ into the
even and odd parts. This allows to check that $\gl_4=0$ is
non-degenerate. In order to compute the magnetic Hessian, we check
that it 
is possible to build a decomposition 
$\Omega ^1(G)=d\R^X \oplus K $
which is $Q$ orthogonal and with $K \subset \ker d^\star $.
It is then easy to check that the magnetic Hessian evaluated
on $K$ vanishes. 

\end{exem}
\section{Bipartite graphs}
\label{sec:bipartite}

Let $G=(V,E)$ be a  bipartite graph, with $V=Y\cup Z$ and all edges
have a vertex in $Y$, the other in $Z$.
Let $U$ be the involution on $\R^V$ given by 
$Uf(x)=-f(x)$ if $x\in Y$ and $Uf(x)=f(x)$ if $x\in Z$
and let $B$ be a magnetic field.
Then $UH_BU=-H'_B $ with $H'\in O_G$.
So that $\gl_{|V|}(H_B)=-\gl_1 (H'_B)$.
An hence it follows form the 
diamagnetic inequality that
  $B \ra \gl_{|V|}(H_B)$ has a maximum at $B\equiv 1$.
And hence the Morse index of the Hessian of $B \ra \gl_{|V|}(H_B)$
at $B\equiv 1$ is the dimension of the manifold of magnetic fields
 namely $\beta$.
On the other hand the first eigenfunction $\phi_1$ of
$H'$ is everywhere $>0$ and the number of sign changes of
$U\phi_1$ is $|E|$.
So Berkolaiko's formula for $\gl_{|V|}$ gives 
 $ (|V| -1)+\beta =|E|$. This is the Euler formula.

\section{Link with the Hessian of the determinant}
\label{sec:determinant}

Let us assume that we are in the discrete case and
the eigenvalue we consider is $\gl_n =0$.
Then we have
\[ {\rm det}(H_B)=\gl_n (B) {\rm det}'(H_B)\]
where  ${\rm det}'(H_B)=F(B)$ is the product of the eigenvalues
$\gl_j$ for $j\ne n$.
We have $(-1)^{n-1}F(1)>0$.
Hence the index of $B \ra (-1)^{n-1}{\rm det }(H_B)$ is 
the same as the index of $B \ra \gl_n (B)$.

There is a formula for the characteristic polynomial
of a magnetic Laplacian on graphs due to Robin Forman \cite{Fo}
and reproved by Kenyon \cite{Ke} and Burman \cite{Bu}.
Using the gauge change $f\ra f \phi_n $ as in my paper 
gives a Laplace type operator whose entries can be of
any sign. Forman's formula extends to that case
and it would be nice to get Berkolaiko's formula form
Forman's formula.


\bibliographystyle{plain}

\end{document}